\documentstyle[11pt,newpasp,twoside,epsf]{article}
\def\edcomment#1{\iffalse\marginpar{\raggedright\sl#1\/}\else\relax\fi}
\reversemarginpar

\begin{document}
\title{The near-infrared  Hubble diagram for sub-mm galaxies}
 \author{James S. Dunlop}
 \affil{Institute for Astronomy, University of Edinburgh,
 Royal Observatory, Edinburgh EH9 3HJ, UK}

\begin{abstract}
Determing the nature of the bright sub-mm sources and their role in the 
process of galaxy formation is likely to be a major focus of 
extra-galactic research over the next few years. In particular, we would
like to know if these sources are the progenitors of today's massive 
elliptical galaxies, or alternatively arise from short-lived, albeit 
spectacular starbursts within haloes of unexceptional mass. This question
can be addressed from a number of different directions, one of which is to 
compare the masses of sub-mm host galaxies with 
those of other known massive high-redshift objects. Here I make a first 
attempt to do this by exploring whether the few well-studied sub-mm/mm 
sources are consistent with the well-established $K-z$ relation for powerful
radio galaxies. Out to $z \simeq 3$ this appears to be the case,
providing evidence that bright sub-mm sources are indeed destined to be 
massive ellipticals. At higher redshifts there is a suggestion that sub-mm
galaxies are signficantly fainter at $K$ than their radio-selected 
counterparts, but at present it is unclear whether this indicates 
a significant difference in stellar mass or the increasing impact of dust 
obscuration on the rest-frame light from the sub-mm hosts.
\end{abstract}

\section{Exploring the nature of the sub-mm galaxy population}

There are various routes by which to address the question of whether or
not sub-mm selected sources are the progenitors of massive elliptical 
galaxies.

One approach is to determine the extent to which known high-redshift 
populations are capable of producing strong mm--far-infrared emission.
This (relatively efficient) experiment has been performed for powerful 
radio galaxies
(Archibald et al. 2001), quasars (e.g. Isaak et al. 2002), 
Lyman-break galaxies (e.g. Chapman
et al. 2000), and 
X-ray selected AGN (e.g. Page et al. 2001). The results of 
these studies indicate that the sort of massive starburst (SFR $\simeq 
500-1000 {\rm M_{\odot} yr^{-1}}$) required to
produce a bright ($S_{850\mu m} > 4$ mJy) sub-mm source is found only 
in the most massive objects, and generally only at high redshift 
($z > 1.5 - 2$). However, it is clearly possible that the sub-mm 
{\it selected} population could be dominated by a somewhat 
different class of source.

A second approach is to compare the estimated comoving number density
of bright sub-mm sources with those of other galaxy populations. As discussed 
by Scott et al. (2002) the comoving number density of sources with 
$S_{850\mu m} > 8$ mJy is $\simeq 1 \times 10^{-5} {\rm Mpc^{-3}}$, comparable to
that of extremely red objects at $z \simeq 1.5$, and present-day 
ellipticals with $L > 2-3 L^{\star}$. However, the interpretation
of such approximate coincidences depends on the assumed typical duration
(and frequency) of the starbursts which power the sub-mm emission.

A third approach is to determine how the clustering properties of the
sub-mm population compare with those displayed by other galaxy populations.
Unfortunately current sub-mm and mm surveys are too small to yield a
statistically significant detection of source clustering.
However, existing data are certainly consistent with clustering as strong 
as that displayed by, for example, extremely red objects (Scott et al. 2002), 
and marginally significant evidence of clustering has been gleaned from 
cross correlation with Chandra sources (Almaini et al. 2002) and 
Lyman-break galaxies (Webb et al. 2002).

A fourth approach, which I explore in this brief article, is to investigate 
how the host galaxies of sub-mm sources compare with other known high-redshift
galaxies. Here I have chosen to make this comparison in the near-infrared
(i.e. $K$-band) both because many of the sub-mm source host galaxies have 
only been detected at $K$, and because observed 
$K$-band brightness currently offers the best means to estimate 
the masses of high-redshift galaxies.

\section{The $K-z$ diagram}

The $K-z$ diagram for radio galaxies has been studied and augmented ever 
since Lilly \& Longair (1984) first demonstrated the existence of a tight 
relation for the 3CR galaxies out to $z \simeq 1$.
Recently Eales et al. (1997), van Breugel et al. (1998) and Jarvis et 
al. (2002) have extended the observed relation to $z > 4$. 
As shown in Fig. 1, and as discussed by 
Jarvis et al. (2002), the radio-galaxy data are consistent with the $K-z$ 
relation expected
for a passively-evolving galaxy of (present-day) luminosity $4-5 L^{\star}$
out to the highest redshifts. Whether pure passive evolution is 
the true explanation of this relation is not really important for the present
purpose. What is clear is that powerful radio galaxies are the most 
massive known galaxies at high redshift and are undoubtedly destined 
to be massive ellipticals (given they contain 
very massive black holes). They thus provide a clear benchmark 
against which to test the hypothesis that sub-mm selected sources are 
the progenitors of today's massive ellipticals.

Before attempting to place sub-mm sources on this diagram it is important
to realize that the $K$ magnitudes of the radio galaxies have all been
measured through large apertures; in Fig. 1 the 3CR and 6C values
have been corrected to a metric aperture of 50 kpc, while the $K$ magnitudes
given by van Breugel et al. (1998) have been measured through an (in effect 
very similar) aperture of diameter 4 arcsec. This is important 
because the continued low scatter in the radio
galaxy relation beyond $z \simeq 2.5$ would not be produced if, for 
example, small aperture magnitudes were simply measured for the 
brightest visible clumps in these often complex sources.
In considering sub-mm sources I have therefore included only
objects with measured or reasonably solid estimated redshifts, and for
which 4-arcsec aperture or near-total $K$ magnitudes have been measured
(or at least attempted) armed with sub-arcsec positional accuracy.

\begin{figure}
\plotone{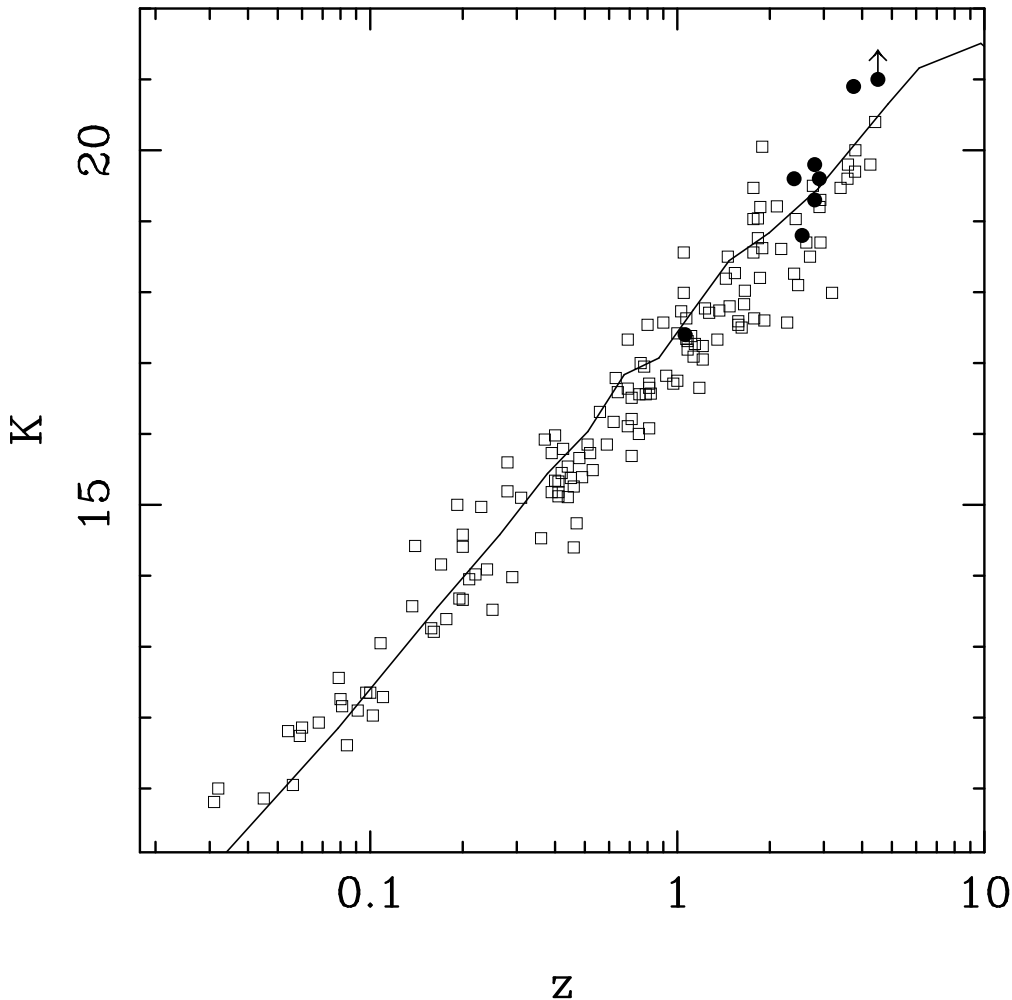}
\caption{\small Evidence that sub-mm sources display a $K-z$
relation similar to that which is well established for radio galaxies.
The open squares indicate the positions of 72 3CR radio galaxies, 57
6C radio galaxies (Eales et al. 1997) and 14 additional high-redshift
radio galaxies (van Breugel et al. 1998) on the $K-z$ plane. The solid line 
is the predicted track of a passively evolving galaxy of constant mass, formed
in an instantaneous burst at high redshift ($z > 10$;
$\Omega_m = 0.3$, $\Omega_{\Lambda} = 0.7$, 
$H_0 = 70 {\rm km s^{-1} Mpc^{-1}}$).  The filled symbols indicate my best estimate
of the locations of eight of the best-studied bright sub-mm sources on the 
$K-z$ plane. The first three of these, from the cluster lens survey
(Smail et al. 2000; Ivison et al. 2000), have been included because they have
spectroscopic redshifts as well as solid measured $K$ 
magnitudes (see text). The next 4 objects are the brightest sub-mm sources 
from the
CUDSS and 8-mJy SCUBA surveys (Eales et al. 1999; Scott et al. 2002), 
all of which have very accurate positions
(from IRAM PdB and VLA interferometry) yielding reliable $K$-band 
identifications in deep near-infrared images. 
Redshifts for these sources have been estimated from their  
1.4GHz/350GHz flux-density ratios as described in the text. Finally, the 
lower limit at $z = 4.5$ has been included to indicate 
the average position of the 3 millimetre sources recently
discussed by Dannerbauer et al. (2002).}
\end{figure}

The first three sub-mm sources I have plotted in Fig. 1 are therefore the 
three 
galaxies with spectroscopic redshifts from the cluster lens survey, namely
SMMJ02399-0134 ($z = 1.06$), 14011+0252 ($z = 2.55$), and 02399-0136 
($z = 2.80$)
(Ivison et al. 2000, Smail et al. 2000). The 
$K$ magnitudes of these three lensed sources have been de-magnified
by factors of 2.5, 2.8 and 2.5 respectively.
To explore the relation further I have added the 4 best-studied 
bright sub-mm sources 
from the
CUDSS and 8-mJy SCUBA surveys (Eales et al. 1999; Scott et al. 2002), 
all of which have very accurate positions
(from IRAM PdB and VLA interferometry) yielding reliable $K$-band 
identifications in deep near-infrared images. 
These are 
CUDSS14A (Gear et al. 2000), Lockman850.1 (Lutz et al. 2001), ELAISN2850.1
and ELAISN2850.2 (Ivison et al. in prep). The redshifts of these 
objects have been estimated from the mean relation between 350GHz/1.4GHz 
flux-density ratio and $z$ given by Carilli \& Yun (2000) which, 
in the redshift range of interest, is effectively identical to the  
extrapolated relation for Arp 220.
Finally, I have included a lower limit at $z = 4.5$
to indicate 
the average position of the 3 millimetre sources recently
discussed by Dannerbauer et al. (2002). These sources merit inclusion because,
while they still lack $K$-band identifications, they 
again have positions from mm-wave interferometry of sufficient accuracy to be 
sure that their counterparts lie below the detection threshold of the
existing $K$-band imaging (to convert the published 2-arcsec diameter 
detection limits to 4-arcsec values I have simply doubled the  
noise).

Despite the obvious uncertainties it can be seen from Fig. 1 that the sub-mm 
galaxies display a plausible relationship between $K$  and $z$. 
Moreover this $K-z$ relation appears indistinguishable to that
displayed by the radio galaxies, at least out to $z \simeq 3$.
At higher redshifts there is a suggestion that sub-mm
galaxies are signficantly fainter at $K$ than radio galaxies, 
but at present it is unclear whether this indicates 
a significant difference in stellar mass or the increasing impact of dust 
obscuration on the rest-frame light from the sub-mm hosts.

\end{document}